\documentclass[12pt]{article}
\usepackage{epsfig}

\def\beq{\begin{equation}}
\def\eeq#1{\label{#1}\end{equation}}
\def\eeqn{\end{equation}}

\def\beqa{\begin{eqnarray}}
\def\eeqa#1{\label{#1}\end{eqnarray}}
\def\eeqan{\end{eqnarray}}

\let\bar=\overbar

\def\Dslash{\not{\hbox{\kern-4pt $D$}}}
\def\dslash{\not{\hbox{\kern-2pt $\del$}}}

\def\msb{{\bar{\ssstyle M \kern -1pt S}}}

\newcommand{\invpb}{\,\mathrm{pb}^{-1}}
\newcommand{\invfb}{\,\mathrm{fb}^{-1}}
\newcommand{\mev}{\,\mathrm{MeV}}
\newcommand{\gev}{\,\mathrm{GeV}}
\newcommand{\ds}{D_s}
\newcommand{\fds}{f_{D_s}}
\newcommand{\ichep}{
  contributed to the {\sl 33rd International Conference on High-Energy
  Physics (ICHEP06)}, Moscow, Russia, July 28 - August 2, 2006}
\def\Title#1{\begin{center} {\Large {\bf #1} } \end{center}}

\begin{document}

\Title{Leptonic and Semileptonic $D$-Decays}

\begin{center}{\large \bf Contribution to the proceedings of HQL06,\\
Munich, October 16th-20th 2006}\end{center}

\bigskip\bigskip

%+\addtocontents{toc}{{\it H. Mahlke}}
%+\label{MahlkeStart}

\begin{raggedright}  

{\it Hanna Mahlke \index{Mahlke, H.}\\
Laboratory of Elementary-Particle Physics\\
Cornell University\\
Ithaca, NY 14853\\
USA}
\bigskip\bigskip
\end{raggedright}

\section{Introduction}

\begin{figure}[htb]
\begin{center}
\epsfig{file=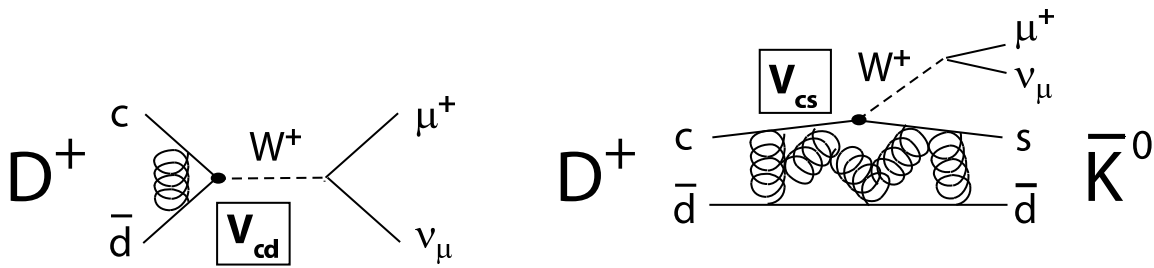,width=1.0\textwidth}
\caption{Diagrams for the reactions $D^+ \to \mu^+ \nu_\mu$ and
$D^+ \to \bar K^0 \mu^+\nu_\mu$.}
\label{fig:processes}
\end{center}
\end{figure}

The study of semileptonic and leptonic decays 
allows the determination of fundamental parameters
of the Standard Model (SM), which are related to
the elements of the
Cabibbo-Kabayashi-Maskawa matrix. 
Examples of these processes are shown in Fig.~\ref{fig:processes}.

The strong interaction of the participating quarks
introduces a complication because its effect is
challenging to estimate accurately.
Theory's difficulty in calculating phenomena rooted in strong
physics seriously hamper some measurements related to weak
physics in the $B$~sector
that need such input, for example extracting 
$V_{ub}$ from semileptonic and leptonic $B$~decays, or 
$V_{td}$ from $B^0-\overline{B^0}$ mixing. 
Theory tools have become available, 
but they require calibration or verification. The similarity
of the charm and the bottom quark allow to test the same methods in
$D$~decay, where the quantities in question can be determined
to excellent accuracy because the corresponding CKM matrix
elements are well-constrained from other sources.
In addition to being of interest in their own right, experimental
charm results thus have a much larger impact on the field.

This presentation discusses leptonic and semileptonic $D$ and
$D_s$~decays,
where results from the BaBar, Belle, CLEO, and FOCUS
collaborations are reported. The challenge in all these
cases, aside from signal purity and background concerns
in general, is to cleanly identify decays with neutrinos
and to determine the kinematic properties of the undetected
particle through external constraints.

The CLEO~experiment benefits from the clean experimental environment
that arises from running at or slightly above production threshold,
$e^+ e^- \to \psi(3770) \to D \bar D$ or
$e^+ e^- \to D^*_s \bar D_s \to \gamma D_s \bar D_s$. 
Identifying one of the $D_{[s]}$ by reconstructing
its decay to a well-identified final state (''tagging'') 
already guarantees a second $D_{[s]}$ in the event,
which is then analyzed for a particular
reaction. Because of the known final state in the first
step, it is also possible to not require a tag, but to
add all reconstructed tracks and showers to infer the neutrino
momentum.
Finally, and of particular importance,
the experimental set-up also allows to count the number of 
produced decays
accurately, thereby facilitating absolute normalization.

The $B$~factories usually use 
the continuum process $e^+e^- \to c \bar c$ at center-of-mass
energies near the $\Upsilon(4S)$. They obtain
$D$~mesons via the decay chain $D^* \to \pi_s D$,
where $\pi_s$ is a slow pion to mark the $D$~flavor. 
It is possible to analyze both $D^+$ and $D^0$ in this way,
but in practice, this method is mostly applied to 
$D^{*+} \to \pi_s^+ D^0$, which also has the most favorable
branching fraction. 
Since determining the number of $D$~decays produced in
fragmentation
is difficult, measurements are often performed relative to
another $D_{[s]}$~decay mode.
Tagging in this case means that the presence of one
reconstructed $D$~meson from $e^+e^- \to c \bar c$ does imply
a second charm particle in the event (which then is
the signal side). Even in the absence of a way to normalize the
rate, it is still possible to compare the shape of 
kinematic distributions with theoretical predictions,
for example form factor shapes (see below).

\section{Leptonic Decays}
Leptonic $D$ decays proceed through annihilation of the constituent
quarks into a $W$, followed by its decay into a lepton and the
corresponding neutrino. 
The partial width for $\ell=e,\mu$, and $\tau$ is given by
\begin{equation}
\Gamma (D^+ \to \ell^+ \nu_\ell) 
 = \frac{1}{8\pi} G_F^2 
f_D^2 m_\ell^2 M_D 
\left( 1 - \frac{m_\ell^2}{M_D^2}\right)^2 |V_{cd}|^2,
\end{equation}
where $f_D$ is the $D$~meson decay constant and
$m_\ell$ and $M_D$ are the lepton and $D$~meson masses.
$G_F$ is the Fermi coupling constant.
For $D_s$ decay, the mass of the $D_s$, the decay constant
of $f_{\ds}$, and $V_{cs}$ are used instead. A similar formula
applies for $B_{[s]}$ decays.
The quantities determined in leptonic $D_{[s]}$ decays are the 
charm decay constant, given that the masses and the CKM
elements are well-measured. The decay constant describes
the strong interaction surrounding the annihilation process.
Experimental accuracy in the few-percent range is crucial
to validate calculations of the decay constant in both
the charm and the bottom sector. Further goals are: comparing
the lepton species to put limits on physics beyond the
Standard Model, and determining $f_{\ds}/f_D$, 
where some technical uncertainties
cancel. The Standard Model predicts the following ratios
for the production rate of $\tau : \mu : e$:
\begin{equation}
\label{eqn:DtoellnuRatios}
D   \to \ell \nu_\ell: \quad 9.72 : 1 : 0.00002, \qquad
D_s \to \ell \nu_\ell: \quad 2.65 : 1 : 0.00002. 
\end{equation}
 
In principle, it is possible to determine the decay constants
from decays to all three lepton species. A few experimental
considerations to take into account:
$D_{[s]}^+ \to e^+ \nu_e$ is beyond current experimental reach 
(low branching fraction due to helicity suppression). 
$D_{[s]}^+ \to \tau^+ \nu_\tau$ is most copiously produced, but
the observed rate is lowered, and the measurement complicated
by, the subsequent decay of the $\tau$, which involves 
at least one more neutrino. 
$D_{[s]}^+ \to \mu^+ \nu_\mu$ is therefore the experimentally
accessible.

\subsection{Leptonic $D$ decays}
\begin{figure}[htb]
\begin{center}
\epsfig{file=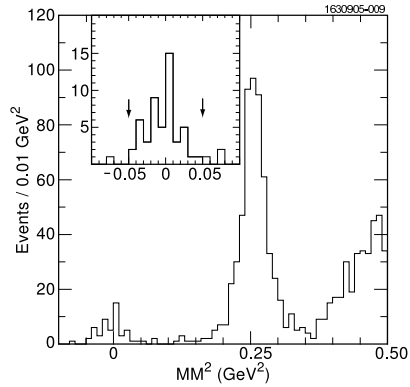,height=2.0in}
\caption{$D^+ \to \mu^+ \nu_\mu$: reconstructed neutrino
mass for signal events~\cite{CLEO:Dtomunu}.}
\label{fig:Dtomunu}
\end{center}
\end{figure}

% ==============================
CLEO studied the decay $D^+ \to \mu^+ \nu_\mu$ in $281\invpb$ of
$\psi(3770)$ data using the tagging 
technique~\cite{CLEO:Dtomunu}. 
The decay is identified with a single muon-like track on the
signal side, and candidate events are required to have a 
missing mass squared near zero.
The missing mass squared is calculated using the beam energy
$E_{\mathrm{beam}}$, the muon energy and momentum
$E_{\mu^+}$ and $p_{\mu^+}$, and the momentum $p_D$ of the $D$: 
$MM^2 = (E_{\mathrm{beam}} - E_{\mu^+})^2 - (-p_{D^-} - p_{\mu^+})^2$,
which for signal events corresponds to the neutrino
mass. 
The signal distribution is shown in Fig.~\ref{fig:Dtomunu}. 
Fifty signal candidates are found, with 
2.8~background events (mostly $D^+$ decay to $\pi^+\pi^0$,
$\tau^+(\to \pi^+\nu)$, and $K^0\pi^+$ tails from the
well-separated peak at higher $MM^2$) expected. This allows measurement of
the branching fraction 
${\cal B}(D^+ \to \mu^+ \nu_\mu) = (4.4 \pm 0.7 \pm 0.1) \times 10^{-4}$ 
and hence $f_D=(222.6\pm16.7 ^{+2.8}_{-3.4})\mev$, 
with $V_{cd}$ as an external input.
With the SM ratios from Eqn.~\ref{eqn:DtoellnuRatios}
applied to the measured branching fraction
${\cal B}(D^+ \to \mu^+ \nu_\mu)$, one expects
${\cal B}(D^+ \to e^+ \nu_e) \sim 1\times 10^{-8}$ and
${\cal B}(D^+ \to \tau^+ \nu_\tau) \sim 2\times 10^{-3}$.
Slight modification of the signal side selection criteria
gives access to $\ell=e, \ \tau$, 
by asking for either an electron-like track (no events
seen, ${\cal B}(D^+ \to e^+ \nu_e) <2.4 \times 10^{-5}$), 
or by asking for
a pion track and a missing mass shifted away from zero,
signalling the decay $\tau \to \pi \nu$ (no significant
signal, ${\cal B}(D^+ \to \tau^+ \nu_\tau) <3.1 \times 10^{-3}$). 

\subsection{Leptonic $\ds$ decays}
\begin{figure}[htb]
\begin{center}
\epsfig{file=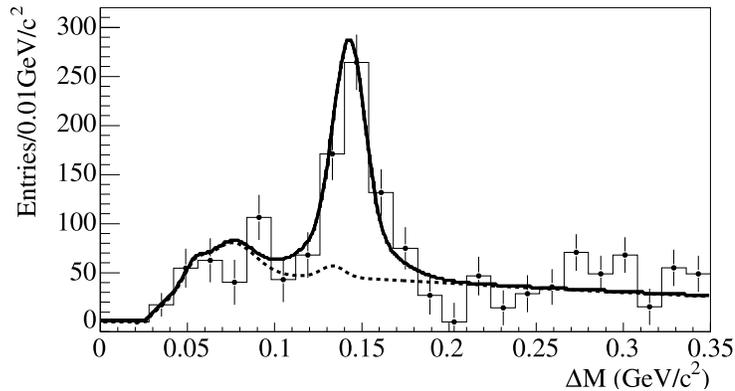,height=2.0in}
\caption{BaBar's signal distribution $\Delta M = m(\gamma\mu^+ \nu_\mu) -
m(\mu^+ \nu_\mu)$ for the decay $\ds^+ \to \mu^+ \nu_\mu$ detected in 
$D_s^* \to \gamma \ds$~\cite{BaBar:Dstomunu}.
The structure at $\Delta M \sim 0.07 \mev/c^2$ is due to the decay
$D_s^* \to \pi^0 \ds$; the peak underneath the signal
in the background shape is due to $\ds^+ \to \tau \nu_\tau$.}
\label{fig:BaBarDstomunu}
\end{center}
\end{figure}

% ================================
Both BaBar and CLEO have recently studied leptonic $\ds$~decays.
BaBar bases their study on $230\invfb$ of data in the 
$\Upsilon(4S)$~region. They use kinematic constraints, 
require an identified $D^0$, $D^+$, $D^{*+}$, or $\ds^+$, 
and use
the signal side decay chain $D_s^{*+} \to \gamma D_s^+$,
$D_s^+ \to \mu^+ \nu_\mu$ to provide a clean $\ds$ 
sample~\cite{BaBar:Dstomunu} (extra particles may be
present in the event). 
The neutrino momentum is inferred from all other 
measured reaction products.
The signal distribution is the $D_s^* - D_s$ mass difference 
$m(\gamma\mu^+ \nu_\mu) - m(\mu^+ \nu_\mu)$, 
shown in Fig.~\ref{fig:BaBarDstomunu}. 
Since the $D_s^*$ production rate is not precisely known,
BaBar determine the ratio relative to $\ds^+ \to 
(K^+K^-) \pi^+$ with $m(K^+K^-)$ in the $\phi$ mass
region ($\pm 2\Gamma_\phi$),
$\Gamma(\ds^+ \to \mu^+ \nu_\mu) / \Gamma(\ds^+ \to \phi \pi^+)$.
With ${\cal B}(\ds^+ \to \phi \pi^+) = 
(4.71 \pm 0.46)\%$~\cite{BaBar:Dstophipi} 
they arrive at ${\cal B}(\ds^+ \to  \mu^+ \nu_\mu) = 
(6.74 \pm 0.83 \pm 0.26 \pm 0.66)\times 10^{-3}$, 
where the third error
is due to the normalization uncertainty. This leads to
$\fds = (283 \pm 17 \pm 7 \pm 14)\mev$. 

CLEO's analysis relies on $\sim 200\invpb$ of data taken
at a center-of-mass energy of $4170\mev$. The decay chain
is $e^+ e^- \to D_s^* \bar D_s \to (\gamma D_s) \bar D_s$.
One $\ds$ is tagged, the other is analyzed for the signal
decay. Two strategies are used. 

A. As above, the missing mass squared is examined to
identify the signal.
The lone signal side track is 
required to not be an electron but instead
to be muonic or pionic. 
An energy requirement of $300\mev$ accepts $99\%$ 
of all muon tracks and $60\%$ of all pion tracks.
We distinguish cases 
with an energy deposition 
below $300\mev$ in the calorimeter (typical for 
$\ds^+ \to \mu^+ \nu_\mu$, but possible for $\tau \to \pi \nu_\tau$) 
from those that have above $300\mev$ (in which case the track is
likely to be a pion from $\tau \to \pi \nu_\tau$).
The data show a clear enhancement in the expected
$\ds^+ \to \mu^+ \nu_\mu$ signal region. The $\ds^+ \to \tau\nu_\tau$
events are more spread out due to the presence of the
additional, unreconstructed, neutrino, but signal events
are seen as well.
The signatures for $\ell =\mu$ and $\tau$ overlap, hence
the summed distribution is fit for the two branching
fractions, where the SM ratio for the two is assumed.
CLEO's preliminary result, combining the two, is
${\cal B} (\ds^+ \to \mu^+ \nu_\mu) = (0.664 \pm 0.076 \pm 0.028)$,
or $\fds = (282 \pm 16 \pm 7)\mev$.
If the track is required to be consistent with an electron,
no candidates are found, resulting in an
upper limit of ${\cal B}(\ds^+ \to e^+ \nu_e) < 3.1 \times 10^{-4}$.

\begin{figure}[htb]
\begin{center}
\epsfig{file=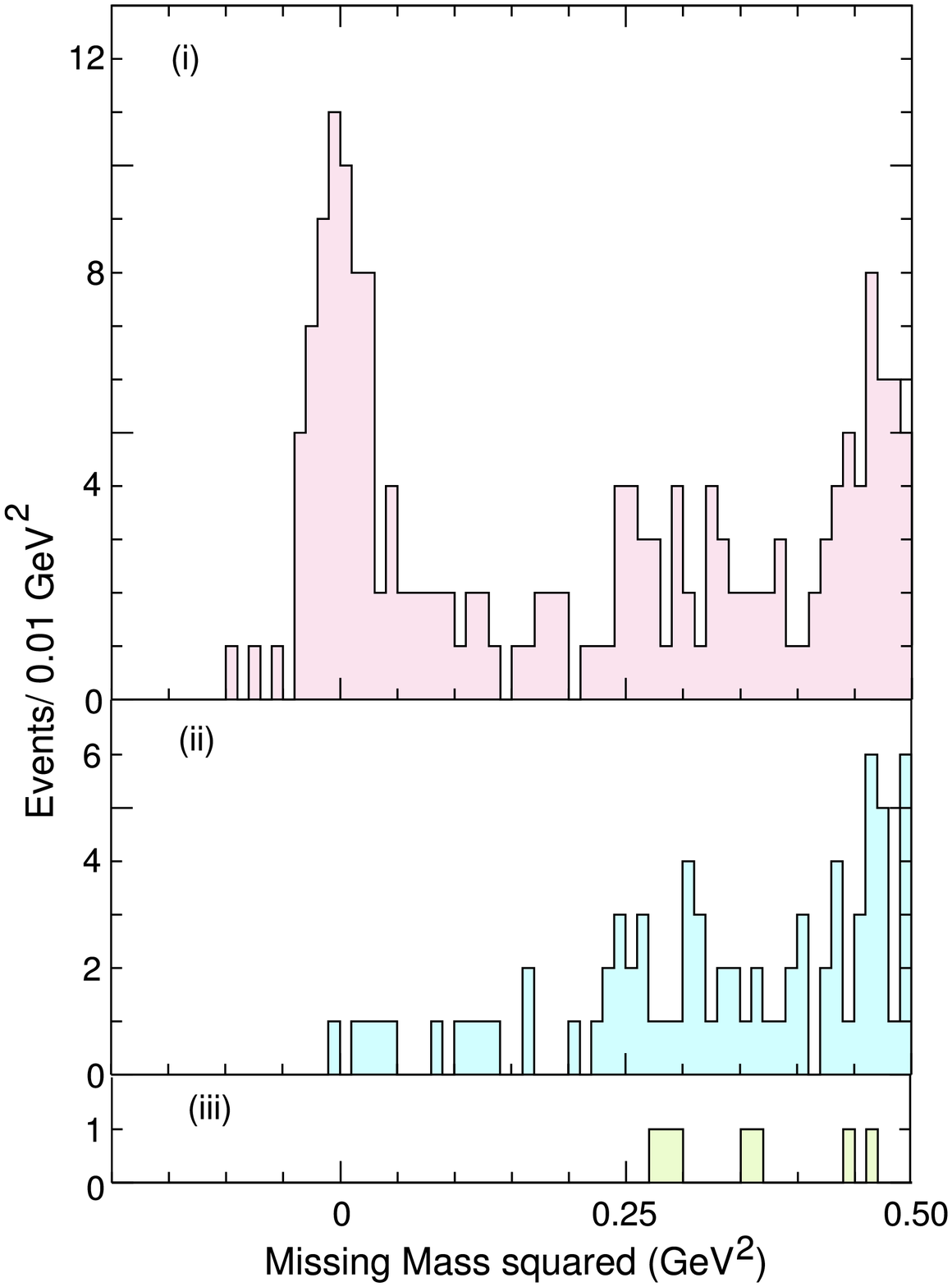,height=3.2in}
\hspace*{4mm}
\epsfig{file=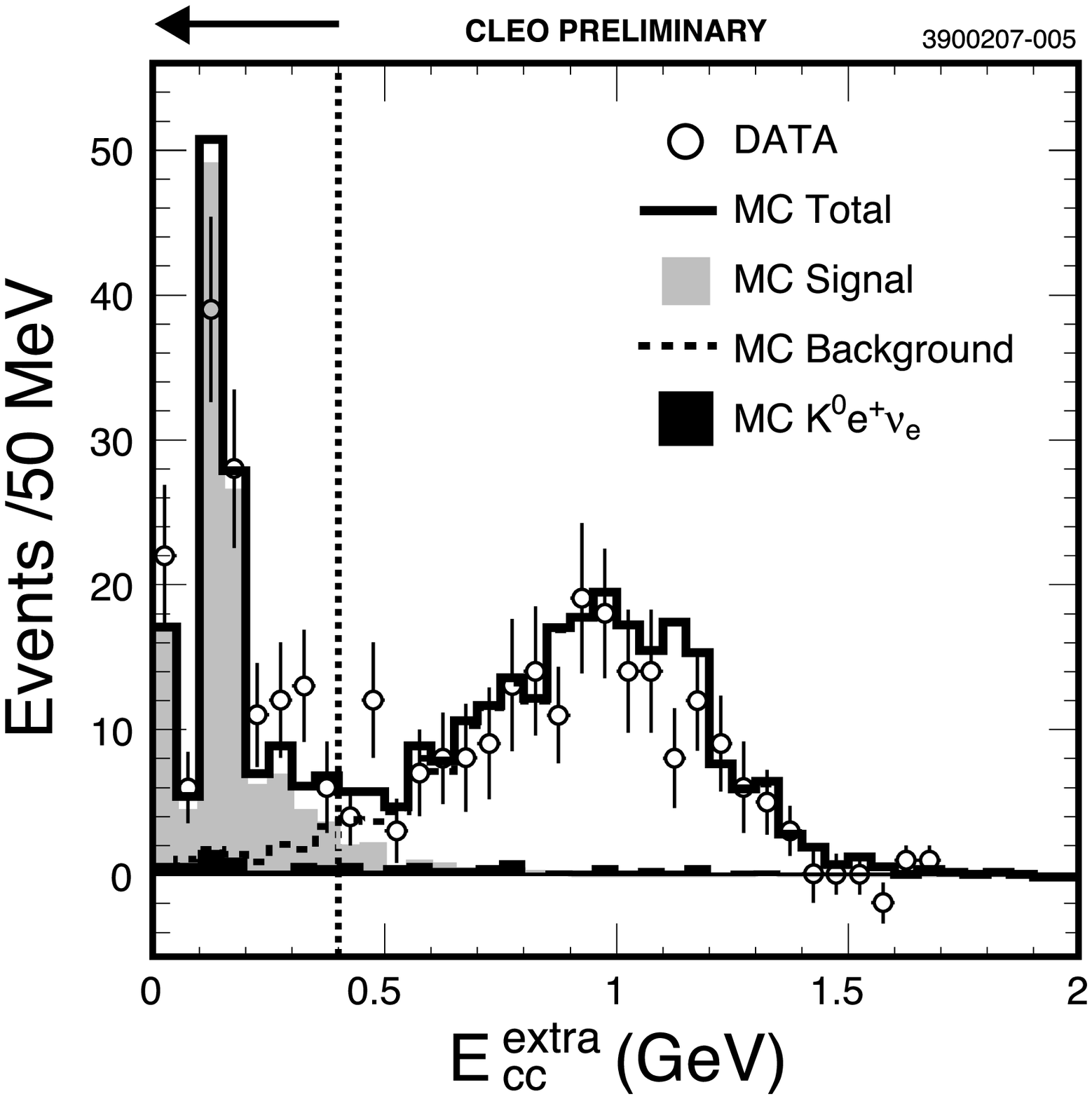,height=3.2in}
\caption{CLEO data on leptonic $D_s$ decays (preliminary).
Left: 
Missing-mass-squared distributions from data corresponding
to (i) $\ds^+ \to \mu^+ \nu_\mu + \tau^+ \nu_\tau$, 
(ii) $\ds^+ \to \tau^+ \nu_\tau$,
(iii) $\ds^+ \to e \nu_e$.
Right:
$\ds^+ \to \tau^+ \nu_\tau$: Energy deposited in the
calorimeter for tagged $\ds$ events and a single
electron-like track on the signal side {\sl not} accounted
for by the tag or the electron candidate. The circles are
data; the curves are MC predictions for signal as well
as several semileptonic background sources.
} \label{fig:CLEODstoellnu}
\end{center}
\end{figure}

B. The second approach, based on the same data, uses
the decay chain $\tau \to e \bar \nu_e  \nu_\tau$.
Its product with the $\ds^+ \to \tau^+ \nu_\tau$ branching
fraction ($\sim 6-7\%$) is about $1.3\%$, to be compared
with the inclusive semileptonic branching ratio
$\ds^+ \to X e^+ \nu_e \sim 8\%$. 
The analysis procedure demands
a sole electron-like track on the signal side, and 
limits the energy not associated with the other
identified decay products in the calorimeter to 
be less than $400\mev$. 
No additional energy deposition is expected for signal 
events other than the transition photon from
$D_s^* \bar D_s \to (\gamma D_s) \bar D_s$,
upon which no selection requirements are placed,
and showers resulting from
interactions of the decay products of the tag side
with the detector material. The signal distribution,
together with background estimates from Monte Carlo simulations,
is presented in Fig.~\ref{fig:CLEODstoellnu}.
This analysis leads to 
${\cal B} (\ds^+ \to \tau^+ \nu_\tau) = (6.3 \pm 0.8 \pm 0.5)\%$
and 
$\fds = (278 \pm 17 \pm 12)\mev$ (both preliminary).
Since the two measurements are complementary, one can form
the average
$\fds = (280 \pm 12 \pm 6)\mev$
and also use them to measure the ratio 
${\cal B} (\ds^+ \to \tau^+ \nu_\tau) : {\cal B} (\ds^+ \to \mu^+\nu_\mu)
=(9.9 \pm 1.9)$, consistent with the SM expectation of 9.72. 

\pagebreak

\subsection{Leptonic Decays: Summary}
\begin{figure}[htb]
\begin{center}
\epsfig{file=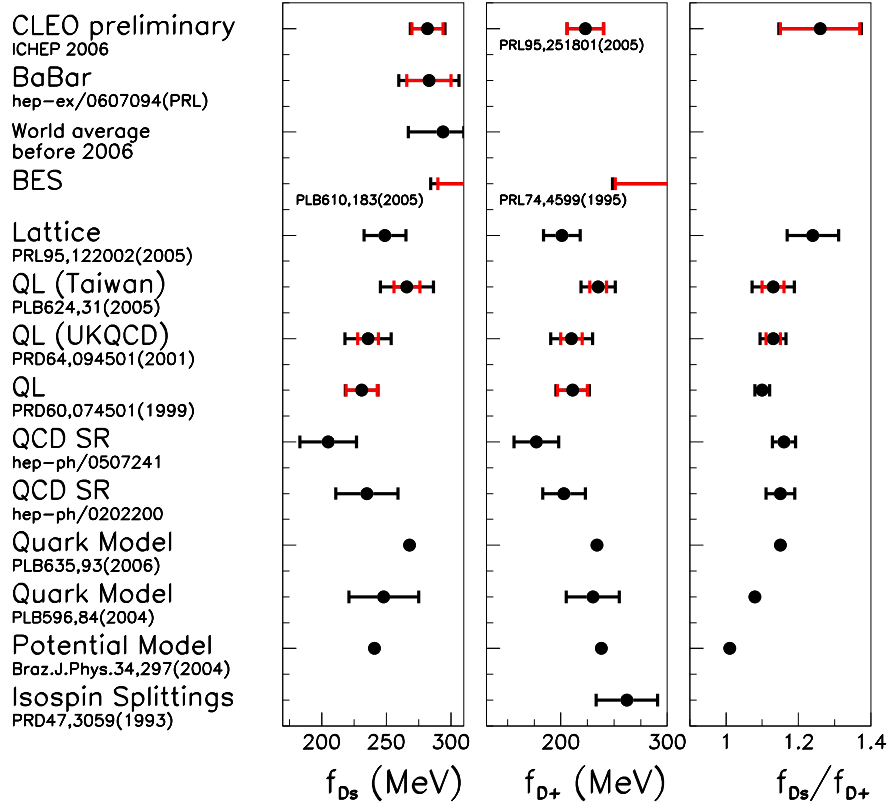, height=4.5in}
\caption{Recent experimental and theoretical results on
$f_D$ and $\fds$.}
\label{fig:fdandfds}
\end{center}
\end{figure}

A visual comparison of current experimental and theoretical
progress on the decay constants is given in 
Fig.~\ref{fig:fdandfds}. In summary: 
$f_D$ from $D^+ \to \mu^+ \nu_\mu$
is measured to a total relative error of 8\%; 
the decays to $e$ and $\tau$ are presently 
not experimentally accessible, although the current
upper limit gives rise to the hope that a signal
will soon be within reach for $\tau$. The decay constant
$\fds$ has been measured in $\ds^+ \to \mu^+ \nu_\mu$ (BaBar, CLEO)
and $\ds^+ \to \tau^+ \nu_\tau$ (CLEO) to an accuracy of 5-8\%.
The precision of theoretical calculations are in this region 
as well. However, the dominant experimental uncertainty
is still statistical, which is -- in principle --
much easier to improve upon than systematic errors such
as those theory is in the process of overcoming.

\pagebreak

\section{Semileptonic Decays}
% ===========================
Semileptonic decays $D \to h \ell \nu_\ell$ provide 
yet another scenario within which to study the impact 
of strong force. The underlying weak process
at the quark level, for instance $c \to W^* q$ with
$W^* \to e \nu$, can be calculated, but the
observed rate is modified by the QCD interaction
between the participant as well as the spectator quarks,
which clouds the simple picture.

\subsection{Branching fractions}
%-------------------------------

The current experimental uncertainty on the $D \to K,\pi e \nu$
branching fractions 
is in the percent regime, thereby posing a challenge
to the precision of the LQCD predictions~\cite{LQCD:DSLX_f0}.

The search is on for rarer modes, and CLEO have improved
considerably upon previously achieved accuracy
in many modes (Table~\ref{tab:DSLX_BR}).
$D^+ \to \eta e^+ \nu$ has been observed for the first
time ($>5\sigma$),
as has the first multi-body semileptonic decay:
$D^0 \to K^- \pi^+\pi^- e^+ \nu_e$ ($>4\sigma$), 
which is found to
mostly proceed via $K_1(1270)^- \to  K^- \pi^+\pi^-$.
The uncertainties on $D^+ \to \omega e^+ \nu$ branching
fraction have been halved, and the upper limits on 
the branching fractions $D^+ \to \eta', \phi e^+\nu$
have been tightened by two orders of magnitude.
Once an $\eta' e \nu_e$ signal is observed this will
allow a comparison with predictions for the ratio
$D \to \eta : D \to \eta'$.

\begin{table}[hb]
\begin{center}
\begin{tabular}{l|ccc}  
\hline
\hline
\rule{0mm}{5mm}
Decay                         &  result ($10^{-2}$)         
                              &  PDG06 ($10^{-2}$)  &  PDG04 ($10^{-2}$) \\
\hline
\rule{0mm}{5mm}
$D^+ \to \bar K^0 \ell^+ \nu$ & T: $8.86 \pm 0.17 \pm 0.20$ 
                              & $8.9 \pm 0.4$       & $6.8 \pm 0.8$\\
                              & U: $8.75 \pm 0.13 \pm 0.30$ 
                              &                     &\\
$D^0 \to K^- \ell^+ \nu$      & T: $3.58 \pm 0.05 \pm 0.05$ 
                              & $3.41 \pm 0.09$     & $3.43 \pm 0.14$  \\
                              & U: $3.56 \pm 0.03 \pm 0.11$ 
                              &                     &\\
                              & Belle: $3.45 \pm 0.07 \pm 0.20$ 
                              &                     &\\
$D^+ \to \pi^0 \ell^+ \nu$    & T: $0.397\pm 0.027 \pm 0.028$ 
                              & $0.44 \pm 0.07$     & $0.31 \pm 0.15$ \\
                              & U: $0.383\pm 0.025 \pm 0.016$ 
                              &                     & \\
$D^0 \to \pi^+ \ell^+ \nu$    & T: $0.309\pm 0.012 \pm 0.006$ 
                              & $0.27 \pm 0.02$     & $0.36 \pm 0.06$\\
                              & U: $0.301\pm 0.011 \pm 0.010$ 
                              &                     & \\
                              & Belle: $0.255\pm 0.019 \pm 0.016$ 
                              &                     & \\
$D^+ \to \rho^0 \ell \nu$     & T: $0.232\pm 0.020 \pm 0.012$ 
                              & $0.24 \pm 0.04$     & $0.31 \pm 0.06$ \\
$D^0 \to \rho^+ \ell \nu$     & T: $0.156\pm 0.016 \pm 0.009$ 
                              & $0.19 \pm 0.04$     & -- \\
\hline
\rule{0mm}{4.5mm}
Decay                            &  CLEO result ($10^{-4}$)         
                                 &  PDG06 ($10^{-4}$)  &  PDG04 ($10^{-4}$) \\
\hline
\rule{0mm}{4.5mm}
$D^+ \to \omega \ell^+ \nu$         & $14.9 \pm 2.7 \pm 0.5$ 
                                 & $16^{+7}_{-6}$      & -- \\
$D^+ \to \phi \ell^+ \nu$           & $ < 2$ (90\% CL)       
                                 & $<209$              & $<209$\\
$D^+ \to \eta \ell^+ \nu$           & $12.9 \pm 1.9 \pm 0.7$ 
                                 & $<70$               & $<50$ \\
$D^+ \to \eta' \ell^+ \nu$          & $ < 3$ (90\% CL)       
                                 & $<110$              & $<110$ \\
\rule[-0.5mm]{0mm}{5mm}
$D^0 \to K^-\pi^+\pi^- \ell^+ \nu$  & $2.9^{+1.9}_{-1.0} \pm 0.5$ 
                                 & $<12$               &  $<12$ \\
$D^0 \to K_1(1270) \ell^+ \nu$      & $2.2^{+1.4}_{-1.0} \pm 0.2$ 
                                 & --              & -- \\
\hline
\hline
\end{tabular}
\caption{CLEO's preliminary~\cite{CLEO:DSLX} and 
Belle's published~\cite{Belle:Kandpienu}
measurements for $D$ semileptonic branching fractions,
and comparison with PDG06~\cite{Yao:2006px} as well
as PDG04~\cite{Eidelman:2004wy}. The PDG06 results include CLEO's
published numbers based on $56\invpb$ of data taken at the
$\psi(3770)$.
``T'' and ``U'' stand for CLEO's ``tagged'' and ``untagged'' analyses,
respectively, and the corresponding
entries are not to be averaged because of sample overlap.
CLEO's results are all obtained with $\ell = e$, Belle's with
$\ell=e$ or $\mu$, and 
the PDG numbers are averaged over $e$ and $\mu$, where available. 
}
\label{tab:DSLX_BR}
\end{center}
\end{table}

\clearpage

\begin{figure}[htb]
\begin{center}
\epsfig{file=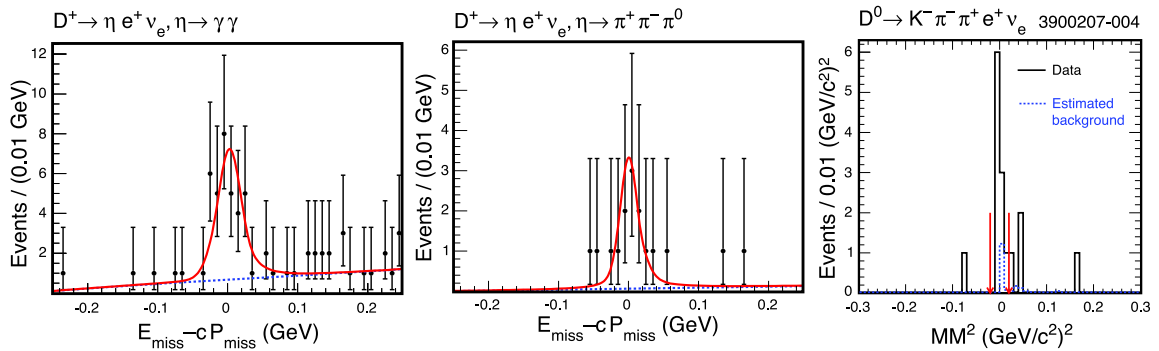,width=1.0\textwidth}
\caption{Preliminary CLEO results: Signal distributions
for $D^+ \to \eta e^+ \nu$ for $\eta \to \gamma\gamma$ (left)
and $\eta \to \pi^+\pi^-\pi^0$ (middle);
$D^0 \to K^- \pi^+ \pi^- e^+ \nu$.}
\label{fig:DSLX_BR}
\end{center}
\end{figure}

\subsection{Form factors}
% =======================

\subsubsection{Decay to a pseudoscalar}
% -------------------------------------
The dynamics of
semileptonic $D \to h \ell \nu_\ell$~decays can for 
the case of a pseudoscalar hadron in the final state
in the limit of small lepton masses
be described as follows:
\begin{equation}
\frac{d\Gamma}{dq^2} \propto
     \left|f_+^h(q^2)\right|^2 \times p_h^3 \times |V_{cq}|^2,
\label{eqn:dsemilep}
\end{equation}
where $q^2$ is the momentum transfer to the $W^*$,
$f_+^h(q^2)$ is the form factor function describing
the probability to end up with a hadron of type $h$
in the final state for a given $q^2$,
$p_h$ is the momentum of the outgoing hadron, and 
$V_{cq}$ is the appropriate CKM~matrix element.

Theoretical predictions for the form factor shape can
be tested against the $q^2$~distribution in data, corrected 
for the $q^2$-dependent detection efficiency. 
This is detailed further below.
The normalization is determined by the product
$|V_{cq}|\times f_+(0)$.
Due to the precision with which $V_{cs}$ and $V_{cd}$
are determined in other experiments ($2-4\%$),
a measurement of $|V_{cq}|\times f_+(0)$ determines
$f_+(0)$.

\begin{figure}[htb]
\begin{center}
\epsfig{file=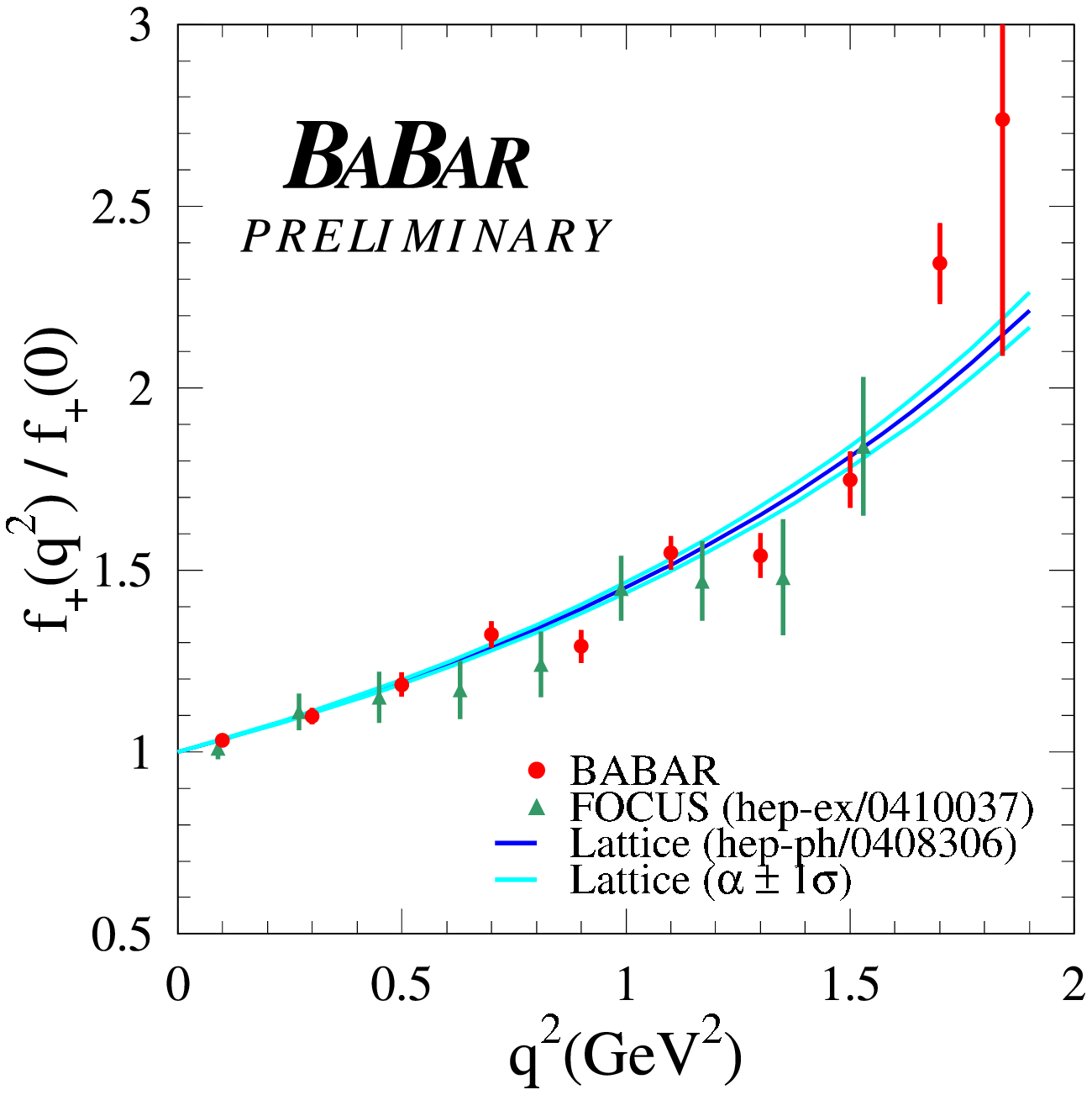,height=2.5in}
\epsfig{file=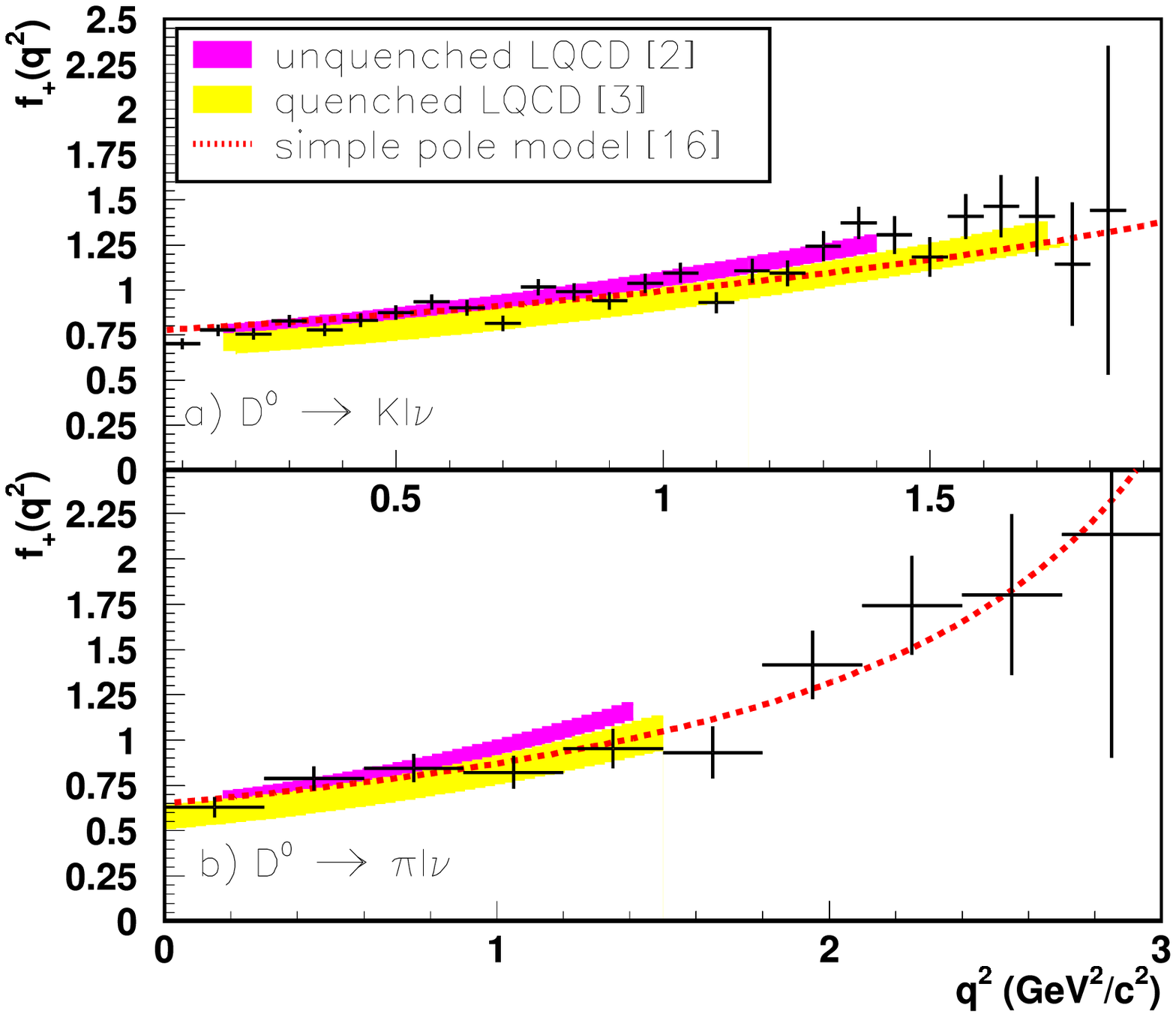,height=2.5in}
\caption{Form factor shapes. 
Left (from Ref.~\cite{BaBar:Kenu}): 
BaBar ($D \to K e \nu_e$, points) and 
FOCUS~\cite{FOCUS:Kmunu} (triangles, $D \to K \mu \nu_\mu$) data 
and an unquenched LQCD calculation~\cite{LQCD:DSLX_f0} (band). 
Right (from Ref.~\cite{Belle:Kandpienu}): 
Belle results on $D \to K$ (top) and $D \to \pi$ (bottom),
with unquenched~\cite{LQCD:DSLX_f0} and quenched~\cite{LQCD:DSLX_quenched} 
LQCD predictions overlaid.
}
\label{fig:ff}
\end{center}
\end{figure}

\begin{figure}[htb]
\begin{center}
\epsfig{file=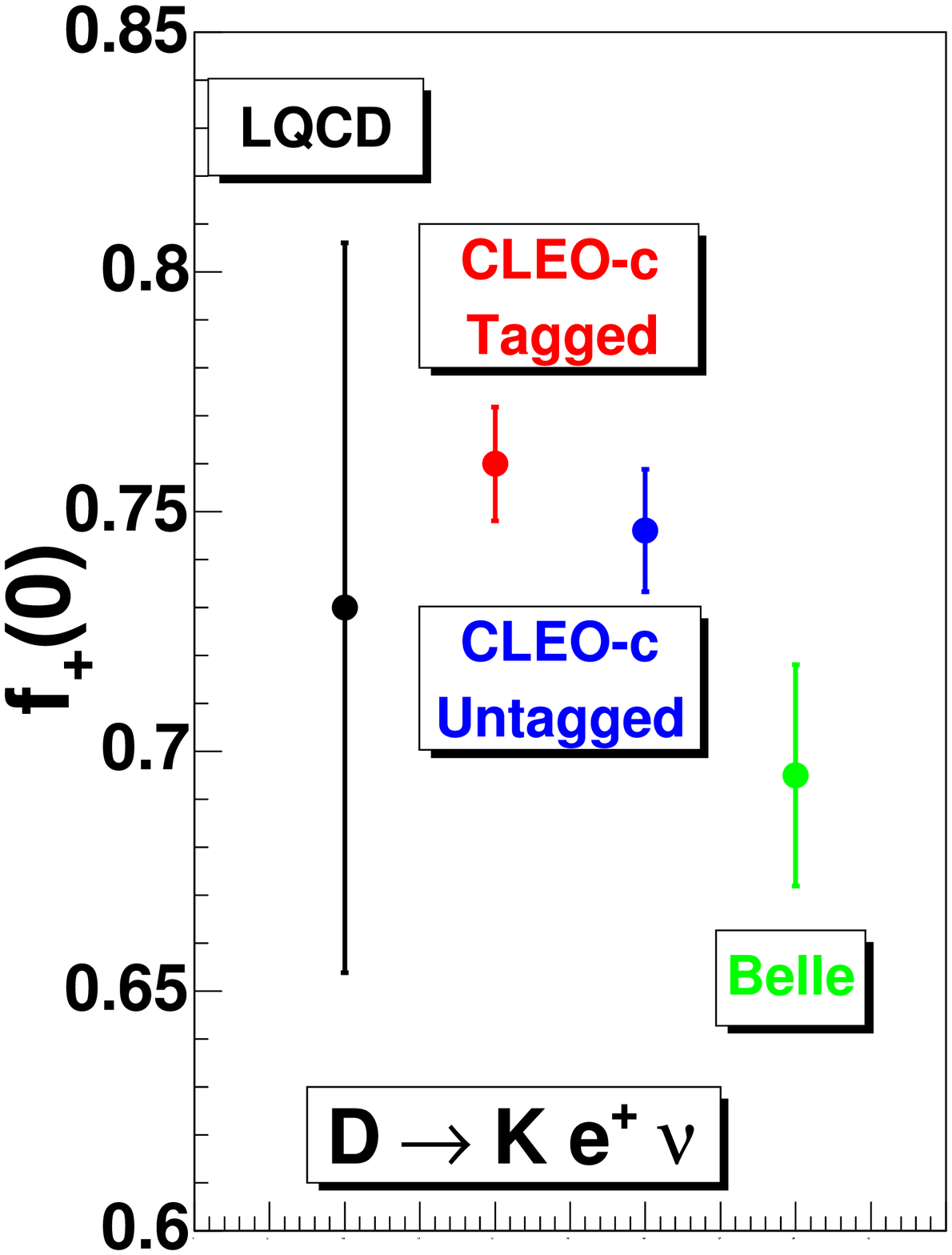,height=2.0in}
\epsfig{file=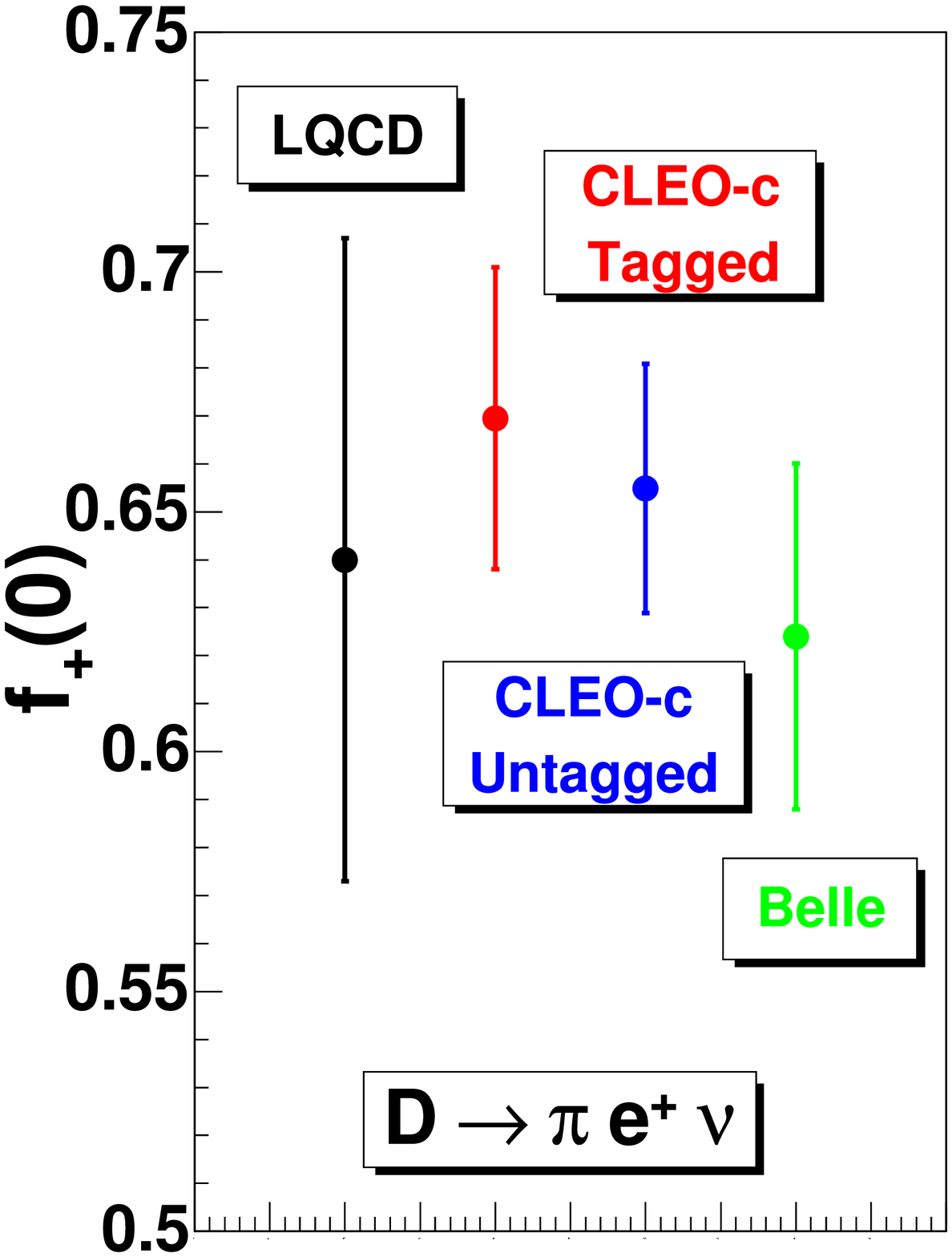,height=2.0in}
\caption{Comparison of form factor normalization for
$D \to K$ and $D \to \pi$ obtained by unquenched LQCD~\cite{LQCD:DSLX_f0},
two CLEO~\cite{CLEO:DSLX} analyses, and Belle~\cite{Belle:Kandpienu}.}
\label{fig:ff_norm_comp}
\end{center}
\end{figure}

Examples of form factor determinations 
are displayed in Fig.~\ref{fig:ff}.
The simplest reasonable parametrization of the 
pseudoscalar form factor is the single pole shape, 
$\sim 1/(1-q^2/M_{\mathrm{pole}}^2)$~\cite{Model:SimplePole},
where the nominal setting of the parameter is
$M_{\mathrm{pole}}=M(D_s^*) [M(D^*)]$ for $h=K[\pi]$.
More sophisticated models are the modified pole model,
\begin{equation}
\label{eqn:modPole}
f_+^h(q^2) = \frac{f_+^h(0)}%
    {(1-\alpha q^2/M_{\mathrm{pole}}^2)\times(1-q^2/M_{\mathrm{pole}}^2)},
\end{equation}
or, recently, the Hill
series parametrization~\cite{Model:Hill}. 
BaBar~\cite{BaBar:Kenu}, Belle~\cite{Belle:Kandpienu},
CLEO~\cite{CLEO:DSLX}, and FOCUS~\cite{FOCUS:Kmunu}
compare their various results for
$D \to \pi,K \ell \nu$ data for $D=D^{+,0}$ with the models
and find reasonable agreement with all of them. 
In particular the Belle and CLEO measurements
boast superb statistics and an excellent $q^2$ resolution.
In general it is found that unquenched LQCD predicts
the form factor shape a little too high especially
for $D\to\pi$~\cite{CLEO:DSLX,Belle:Kandpienu},
but has achieved useful uncertainty, as visually
evident from the agreement between the data points and
the curve obtained from an interpolation between 
LQCD predictions at several $q^2$ points, Fig.~\ref{fig:ff}. 
Fitting the predicted LQCD data points to the modified pole
model shape, thereby extracting $f_+(0)$ and $\alpha$,
allows a comparison within this model between theory
and the experiments. This is presented for $f_+(0)$ in
Fig.~\ref{fig:ff_norm_comp}.

Alternatively, using
$f_+^{D\to\pi} = 0.64 \pm 0.03 \pm 0.06$ and
$f_+^{D\to K} = 0.73 \pm 0.03 \pm 0.07$ 
from LQCD~\cite{LQCD:DSLX_f0}, the CKM~matrix
elements are found to be in good agreement with
current world averages. However, the uncertainties
($\sim 10\%$ relative) on the form factor magnitudes
dominate. This constitutes a check of the LQCD calculation.

\subsubsection{Decay to a vector}
% -------------------------------

For a vector hadron in the final state,
such as $D \to K^* e \nu$ (Cabibbo-favored)
or $D \to \rho e \nu$ (Cabibbo-suppressed), 
more form factor functions enter the stage.
The decay amplitude
can be described in a parameter-free way 
using helicity basis form factors as specified in 
Ref.~\cite{FOCUS:NonParamFF}.  
However, the helicity form factors are hard
to calculate; a more traditional approach is to just
assume spectroscopic pole dominance and cast the
expression for the amplitude
into linear combinations of the following functions:
\begin{equation}
\label{eqn:vecFF}
A_i(q^2) = \frac{A_i(0)}{1-q^2/M_{A_i}^2} (i=1,2), \newline
V(q^2) = \frac{V(0)}{1-q^2/M_{V}^2}.
\end{equation}
The pole masses are often fixed to be $M_V=2.1\gev$,
$M_{A_{1,2}}=2.5\gev$. If one then defines 
$R_V = V(0)/A_1(0)$ and $R_2 = A_2(0)/A_1(0)$,
one ends up with only two free parameters, albeit
at the expense of an assumed shape. The parameter-free
approach in Ref.~\cite{FOCUS:NonParamFF} circumvents this.

\begin{figure}[htb]
\begin{center}
\epsfig{file=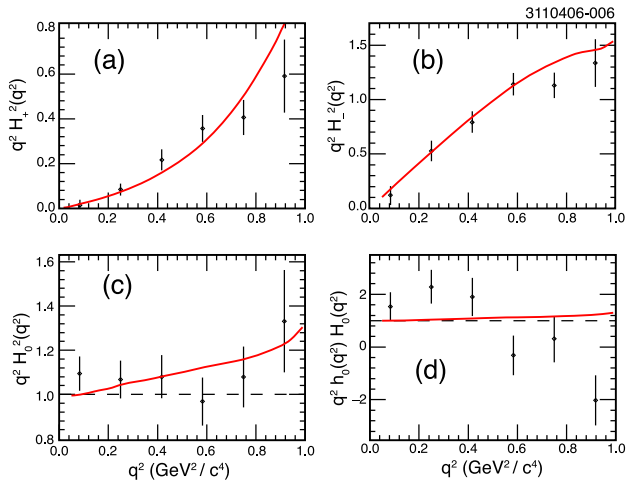,height=2.5in}
\caption{$D \to K\pi e \nu$~\cite{CLEO:NonParamFF}:
Four parameter-free helicity-basis form factor products as 
extracted in CLEO data, overlaid with form factors 
constructed from spectroscopic pole model shapes 
using $R_0 = 1.505$, $R_2 = 0.875$, and $s$-wave parameters
${\cal A} = 0.33$ and $\delta = 39^\circ$.}
\label{fig:NonParamFF}
\end{center}
\end{figure}

An analysis of $D^+ \to K^- \pi^+ e^+\nu$ CLEO data uses
a projective weighting technique, by which the expected
shapes from helicity form factor contributions are fit
to the data, thereby extracting the (bin-wise) amplitude
for each of the form factors~\cite{CLEO:NonParamFF}. When the
spectroscopic form factors as given in Eqn.~\ref{eqn:vecFF},
translated into helicity form factors,
are overlaid, good agreement is found (Fig.~\ref{fig:NonParamFF}).
Additional conclusions are that a term describing interference
with a non-resonant $s$-wave $K\pi$ component is necessary, 
and that no evidence is found for $d$- or $f$-wave contributions.

A recent preliminary study by BaBar~\cite{BaBar:Ds_phienu} 
of $D_s^+ \to \phi e^+ \nu$, using about $13\times 10^3$
signal events in $78.5\invpb$, not only
exhibits a beautifully precise measurement
but also confirms an earlier FOCUS result.
This actually resolves a controversy: The ratios
$R_V$ and $R_2$ for $D^+ \to K^* e^+ \nu$ and
$D_s^+ \to \phi e^+ \nu$ are expected to be similar
because the CKM matrix element involved is $V_{cs}$
in both cases, and the only remaining difference is
then the spectator quark flavor ($\bar d$ {\it vs.}~$\bar s$).
Agreement within 10\% is expected. This was not borne
out by earlier data, which was not inconsistent for $R_V$
but disagreed for $R_2$, 
except for a FOCUS 
measurement~\cite{FOCUS:Ds_phienu} that had $R_2^\phi$ 
in agreement with the world average of $R_2^{K^*}$.

\begin{figure}[htb]
\begin{center}
\epsfig{file=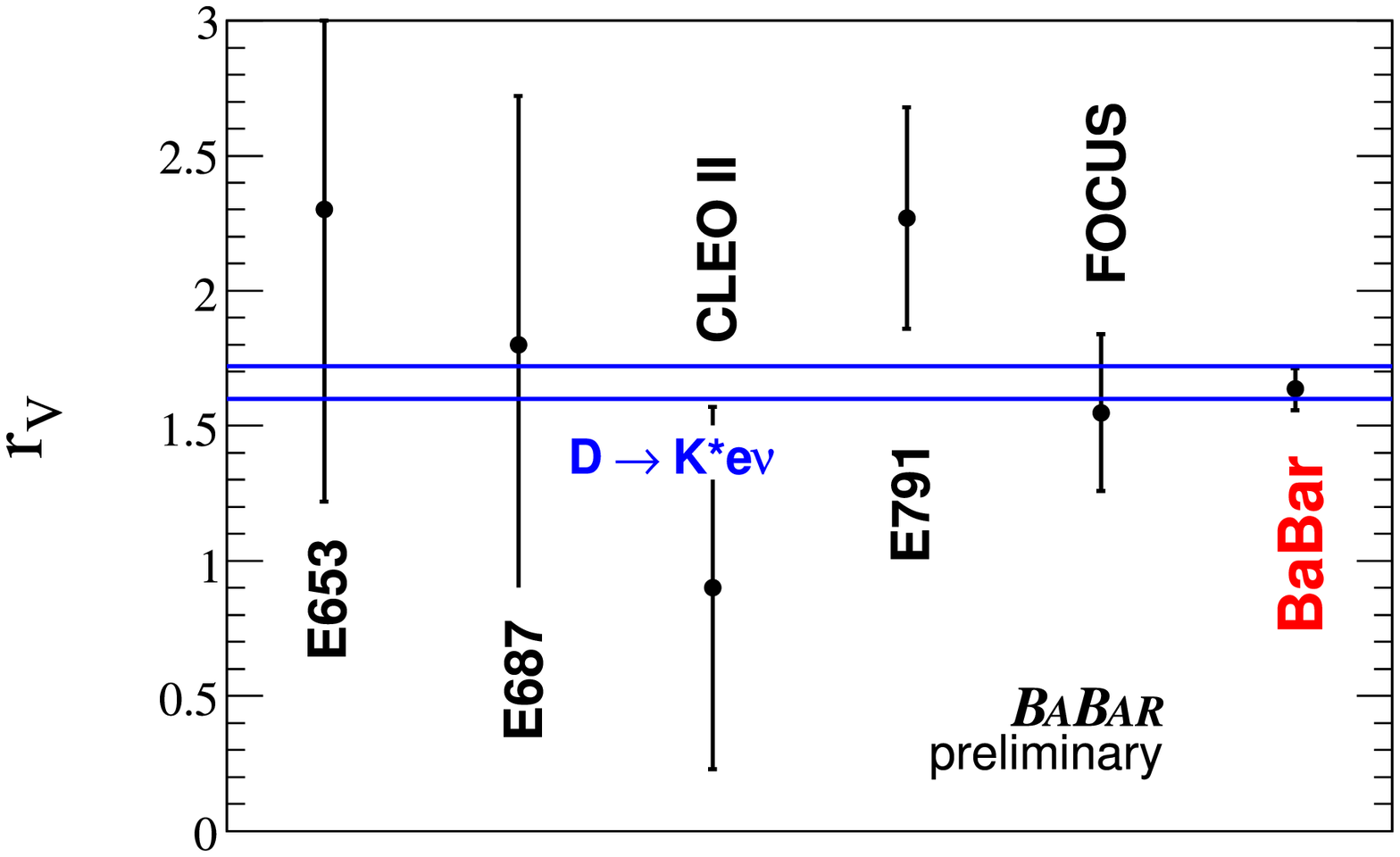,height=2.0in}
\epsfig{file=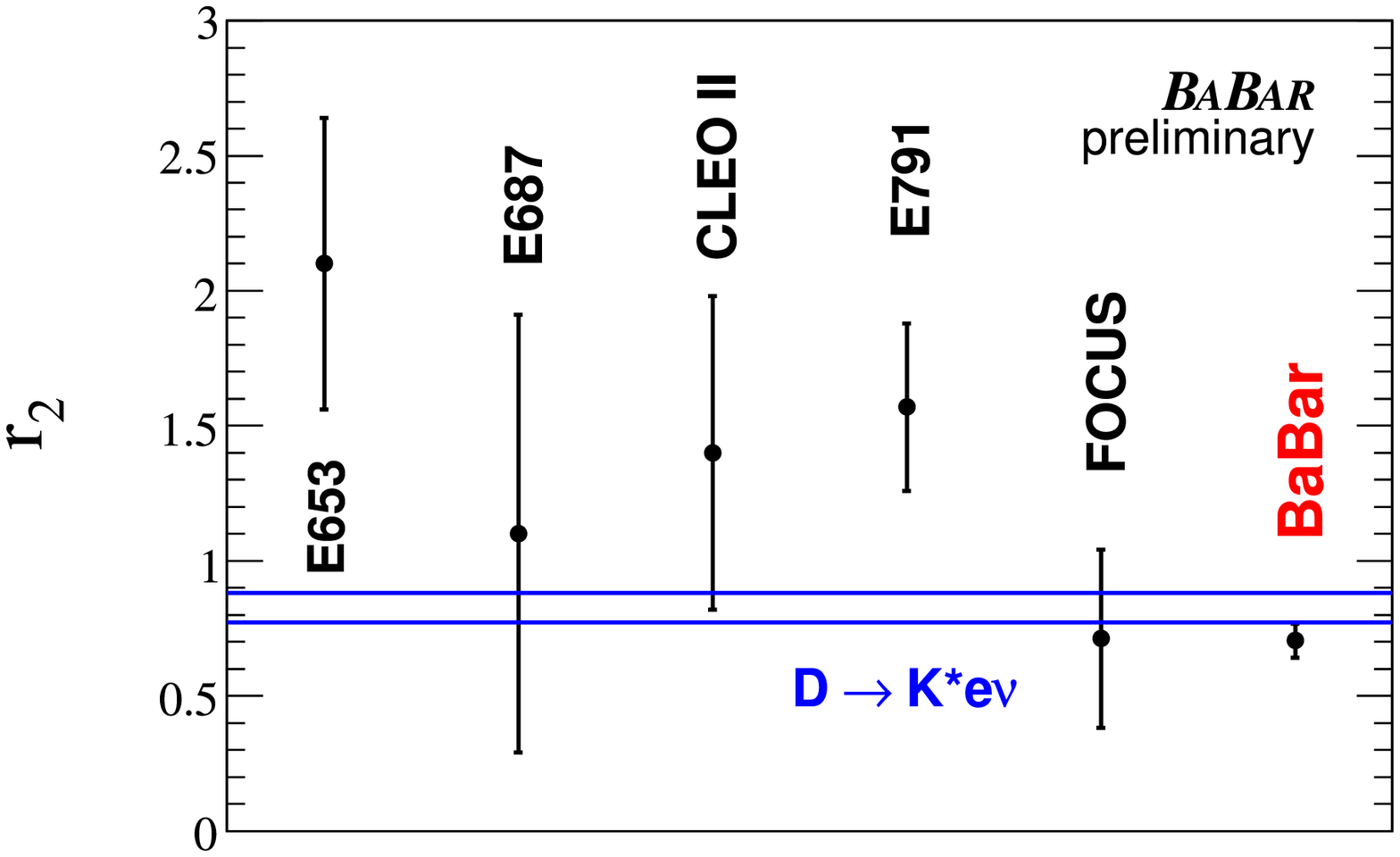,height=2.0in}
\caption{
Comparison of form factor parameters $R_2$ (left)
and $R_V$ (right)
for $D \to K^*$ and $D_s \to \phi$ semileptonic decays,
from Ref.~\cite{BaBar:Ds_phienu}. The data points
all represent $D_s \to \phi$ analyses. The
lines give the $\pm 1\sigma$ band for an average
of experimental $D \to K^*$ results.}
\label{fig:RVR2}
\end{center}
\end{figure}

CLEO has presented preliminary results for the 
first Cabibbo-suppressed 
pseudoscalar $\to$ vector form factor measurement,
$D \to \rho e \nu$. A clear signal for both isospin
states is observed, 
making it possible to bin in kinematic distributions.
The branching fractions from the $281\invpb$~data
sample are found to be in good agreement 
with previous results (which are dominated by the
CLEO-c results from the $56\invpb$~sample,
see Table~\ref{tab:DSLX_BR}), 
as is the partial width 
$D^0 \to \rho^- e^+ \nu = (0.41 \pm 0.03 \pm 0.02) \times 10^{-2}$.
A simulatenous fit to
$D^+ \to \rho^0 e^+\nu_e$ and $D^0 \to \rho^- e^+\nu_e$
(linked through isospin) results in the following
determinations: $R_V^\rho = 1.40 \pm 0.25 \pm 0.03$,
$R_2^\rho = 0.57 \pm 0.18 \pm 0.06$.

$D \to \rho e \nu$ and $D \to K^* e \nu$ studies are important
to the $B$~sector for example by aiding in the
extraction of $|V_{ub}|^2 / |V_{ts}|^2$ through
$d\Gamma(B \to \rho e \nu) : 
d\Gamma(B \to K^* \ell\ell)$~\cite{GrinsteinPirjol},
which requires either measurement of the $B\to\rho$
and $B \to K^*$~form factors or a calculation validated 
by the corresponding $D$~decays.

\subsection{Inclusive semileptonic decays}
% ----------------------------------------
CLEO has determined the semileptonic inclusive branching
fraction $D \to X e^+ \nu_e$ for both charged and neutral
$D$~mesons~\cite{CLEO:DSLIncl}. 
The measured branching fractions
${\cal B}(D^0 \to X e^+ \nu_e)=(6.46 \pm 0.17 \pm 0.13)\%$,
${\cal B}(D^+ \to X e^+ \nu_e)=(16.13 \pm 0.20 \pm 0.33)\%$
agree well with the sum of all exclusive modes, 
although there is room for as-yet-unobserved
exclusive decays at the level of ${\cal B} \sim 10^{-3}$.
Isospin symmetry is observed
within errors for the inclusive semileptonic partial widths,
$\Gamma_{D^+} / \Gamma_{D^0} = 0.985 \pm 0.028 \pm 0.015$.

\begin{figure}[b!]
\begin{center}
\epsfig{file=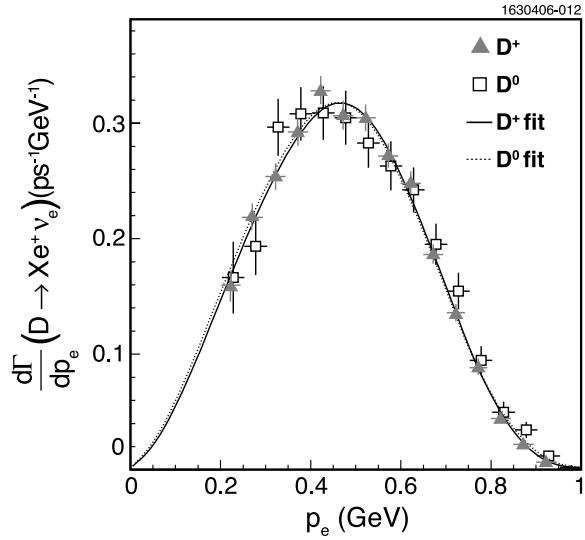,height=2.8in}
\caption{Electron momentum spectrum in $D \to Xe^+\nu_e$
for charged and neutral $D$~mesons~\cite{CLEO:DSLIncl}.}
\label{fig:DSLIncl}
\end{center}
\end{figure}

Aside from the branching fraction, a quantity
of interest is the electron momentum spectrum. Without
weak annihilation contributions ($D^+ \to g W^+ \to X e^+ \nu$)
agreement between the spectra from $D^+$ and $D^0$ is
expected. Weak annihilation would modify the upper end
of the electron spectrum. Within uncertainties, the 
$D^0$ and $D^+$ distributions agree (Fig.~\ref{fig:DSLIncl}).

\subsection{Semileptonic decays: Summary}
% ---------------------------------------
Many new results on $D$ branching fractions and 
$D \to P,V$ and $\ds \to V$ form factors have become
available. Of particular interest 
are the shape and the
normalization of the form factor functions. 
Experimental accuracy is at present still consistent
with most calculations on the market; this
may change as the experimental and theoretical
uncertainties decrease.

\section{Conclusions}
% ===============
Precise results in the $D$~sector have improved our
comprehension of the QCD effects accompanying weak
interactions and allowed to sharpen theoretical tools.
Thanks to the similarity of the heavy quarks $c$ and
$b$, common calculation techniques can be applied to the 
estimation of $D$ and $B$ decay properties.
Further progress in the $D$~sector and consequently
the $B$~sector is in sight as data samples with modern
detectors are being enlarged.

\section{Acknowledgements}
% ========================
\bigskip
I would like to thank the conference organizers for an enjoyable
conference at a exceptionally interesting venue. I am further indebted 
to many of my colleagues on CLEO, BaBar, Belle and FOCUS for helpful
discussions. This work was supported by the 
US National Science Foundation under cooperative agreement PHY-0202078.

\end{document}